\documentclass[english,pre,twocolumn,showpacs,floatfix]{revtex4}
\usepackage{babel}
\usepackage{epsfig}
\usepackage{amsmath}
\usepackage{amssymb}

\DeclareMathOperator{\Trace}{Trace}

\begin{document}
\title{Self-organized escape of oscillator chains in nonlinear potentials}
\author{ D. Hennig,$^{1}$ S. Fugmann,$^{1}$ L.
  Schimansky-Geier,$^{1}$ and P. H\"anggi$^{2}$}

\affiliation{ $^{1}$Institut f\"{u}r Physik, Humboldt-Universit\"{a}t
  Berlin, Newtonstrasse~15, D-12489 Berlin, Germany\\
  $^{2}$Institut f\"{u}r Physik, Universit\"{a}t Augsburg,
  Universit\"{a}tsstrasse~1, D-86135 Augsburg, Germany}

\newpage
\begin{abstract}
  \noindent We present the noise free escape of a chain of linearly interacting units from a
metastable state over a cubic on-site potential barrier. The underlying dynamics is conservative and
purely deterministic.
The mutual interplay  between nonlinearity and harmonic interactions
causes an initially uniform lattice state  to become unstable,
leading to an energy redistribution with strong localization.
As a result a spontaneously emerging localized mode grows into a
critical nucleus. By surpassing this transition state, the nonlinear
chain manages a self-organized, deterministic barrier crossing. Most
strikingly, these noise-free, collective nonlinear escape events
proceed generally  by far faster than transitions assisted by
thermal noise when the ratio between the average energy supplied per
unit in the chain and the potential barrier energy assumes small
values.

\end{abstract}
\maketitle


\section{Introduction}
The cornerstone work by Kramers (for a comprehensive review see
Ref.~\cite{RMP}) has instigated an ever ongoing interest in the
dynamics of escape processes of single particles, of coupled degrees
of freedom or of small chains of coupled objects out of metastable
states. In undergoing an escape the objects considered manage to
overcome an energetic barrier, separating the local potential
minimum from a neighboring attracting domain.

Stochastic---i.e., noise-assisted---escape is the predominant
phenomenon being studied in statistical physics. 
Then, the system energy fluctuates
while permanently interacting with a thermal bath. This causes dissipation 
and local energy fluctuations and the process is
conditioned on the creation of a rare optimal fluctuation which in
turn triggers an escape \cite{RMP}. In other words, an optimal
 fluctuation transfers sufficient energy to the chain so that
the latter statistically overcomes the energetic bottleneck.
Characteristic time-scales of such a  process are determined by the
calculation of corresponding rates of escape out of the domain of
attraction. In this realm, many extensions of Kramers escape theory
and of first passage time problems in over- and underdamped versions
have been widely investigated \cite{RMP,JSPHa}. Early
generalizations to multi-dimensional systems date back to the late
$1960$'s \cite{Langer}. This method, nowadays, is commonly utilized
in biophysical contexts and for a great many applications occurring
in physics and chemistry \cite{Sung}-\cite{BM}.

With this work we present
a different scenario of a possible exit from a metastable domain of
attraction which has  recently been put forward in Ref. \cite{EPL}.
The model we shall study is a {\it purely deterministic} dynamics of
a linearly coupled chain of nonlinear oscillators. Put differently,
no additional coupling to a thermal bath promotes the escape.
Henceforth, dissipation vanishes as well within this set up. The
underlying mechanism to create an escape is caused solely by the
strongly nonlinear Hamiltonian deterministic dynamics.

We explore macroscopic discrete, coupled nonlinear oscillator chains
with up to $1000$ links.  These may appear as realistic models in
mechanical and electrical systems, in various biophysical contexts,
in neuroscience, or in networks of coupled superconductors, to name
but a few  \cite{neuro}-\cite{breathers}. A deterministic escape in
the absence of noise is particularly important in the case of low
temperatures when activated escape becomes far too slow. Also the
case of many coupled nonlinear units in the presence of non-thermal
intrinsic noise that scales inversely with the square root of the
system size then calls, in the limit of large system sizes, for a
deterministic  nonlinear escape scenario.

In the nonlinear regime an initially, almost homogeneous chain is
able to generate spontaneous structural modulations. This process
proceeds in a self-organized manner. More specifically, due to the
modulational instability, unstable growing nonlinear modes give rise
to the formation of coherent structures \cite{Remoissenet} such that
the originally uniformly distributed energy becomes concentrated to
a few degrees of freedom. With regard to nonlinear localization
phenomena {\it intrinsic localized modes or discrete breathers}, such as
time-periodic and spatially localized solutions of nonlinear lattice
systems, have turned out to present the archetype of localized
excitations in numerous physical situations
\cite{MacKay}-\cite{Josephson}.

An escape is related to a crossing of a saddle point in
configuration space, corresponding to a bottleneck \cite{RMP} or a
transition state. The latter is associated with an activation energy
$E_{act}$ to be concentrated in the critical localized mode. We will
show that the critical localized mode can be reached in the
microcanonical situation spontaneously \cite{EPL}. Thus, we
encounter a self-organized creation of the transition state which is
in clear contrast to noise activated escape. In particular, we
demonstrate that intrinsic {\it nonlinear effects} on a long
discrete chain of $N$ units induce a transition over an energetic
barrier by enhancing one, or several localized breather states
\cite{Flach}-\cite{breathers}. Due to this mechanism the initially
almost uniformly distributed energy is dynamically concentrated by
use of an internal redistribution; no assistance of energy exchange
with a thermal bath is thus needed. We show as well that the
nonlinear mechanism of energy localization may promote a {\it
faster} escape dynamics as compared to the noise-assisted situation
where the system experiences an enduring stochastic forcing.

The paper is organized as follows: In the next section we introduce
the model of the coupled oscillator chain and discuss in
Sec.~\ref{section:modulational} the modulational instability as the
localization mechanism.
In Sec.~\ref{section:breather} we proceed by focusing our  interest on the low-energy modes
corresponding to nearly equilibrium states of the lattice chain. The
properties of localization induced by the dynamical formation of
breather arrays are explored. Concerning the escape itself,  special
attention is paid to the passage of lattice states through a
critical localized mode (transition state) in Sec.~\ref{section:transition}.
Subsequently, in Sec.~\ref{section:resonance}
we demonstrate that the rate of escape may be crucially affected by the
coupling strength.
In Sec.~\ref{section:comparison} the escape rates obtained under microcanonical conditions are
compared with those found for thermally activated barrier crossings.
In this context we assume not only flat-state initial preparations
of the microcanonical system but also random chain configurations
with a fairly broad distribution of the coordinates and/or momenta.
In Sec.~\ref{section:length} we deal with the influence of the chain
length on the escape process.  We conclude with a summary of our
results.

\section{Coupled oscillator chain model}

We study a one-dimensional lattice of coupled nonlinear oscillators.
Throughout the following we shall work with dimensionless
parameters, as obtained after appropriate scaling of the
corresponding physical quantities.
The coordinate $q$ of each
individual oscillator of mass unity evolves in a cubic, single well
on-site potential of the form
\begin{equation}
U(q)=\frac{\omega_0^2}{2}q^2-\frac{a}{3} q^3
\,.\label{eq:potential}
\end{equation}
This potential possesses a metastable equilibrium at $q_{min}=0$,
corresponding to the rest energy $E_{min}=0$ and the  maximum is
located at $q_{max}=\omega_0^2/a$ with energy $E_{max}\equiv \Delta E=\omega_0 ^6/(6 a ^2)$.
Thus, in order for particles to escape
from the potential well of depth $\Delta E$ over the energy barrier
and subsequently into the range $q>q_{max}$ a sufficient amount of
energy has to be supplied.

\begin{figure}
  \begin{center}
    \begin{minipage}{8.cm}
      \resizebox{8.cm}{6.cm}{\includegraphics[scale=1.0]{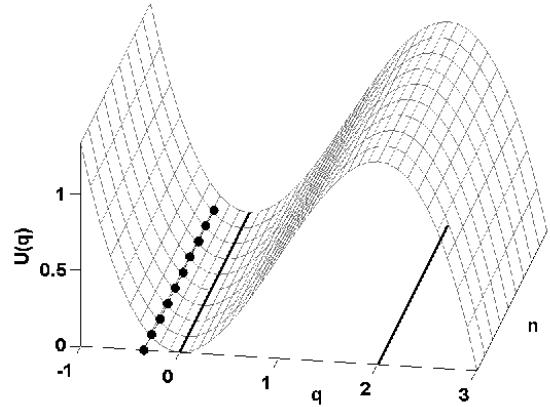}}
    \end{minipage}
    \begin{minipage}{5.cm}
    \end{minipage}
\end{center}
\caption{\label{fig:potential}
 Potential barrier landscape with a chain positioned at the bottom.
The parameter values are $a=1$ and $\omega_0^2=2$.}
\end{figure}

The Hamiltonian of the one-dimensional coupled nonlinear oscillator chain
is given by
\begin{equation}
H= \sum_{n=1}^N\,\left\{\,\frac{p_{n}^{2}}{2}
\,+\,U(q_n)\,\right\}+\,\frac{\kappa}{2}\,\sum_{n=1}^{N}\,\left[\,q_{n+1}-q_{n}\,\right]^2\,.
\label{eq:Hamiltonian}
\end{equation}
The coordinates $q_{n}(t)$ quantify the displacement of the
oscillator in the local on-site potential $U$ at lattice site
$n$ (see Fig.~\ref{fig:potential}),  and $p_n(t)$ denotes the
corresponding canonically conjugate momentum. The oscillators, also
referred to as "units'', are coupled linearly to their neighbors
with interaction strength $\kappa$.


The equations of motion derived from the Hamiltonian given in
Eq.\,(\ref{eq:Hamiltonian}) then read:
\begin{equation}
\frac{d^2q_{n}}{dt^2}\,+\,\omega_0^2 q_n-a q_{n}^2-\kappa\,\left[\,q_{n+1}+q_{n-1}-2q_n\,\right]
=0\,\,.\label{eq:qdot}
\end{equation}
Throughout this work we use periodic boundary conditions according
to $q_{1}(t)=q_{N+1} (t)$.
%
%
Note that in Eq.\,(\ref{eq:qdot}) the
nonlinearity is solely contained in the local force term.

For a setup with interacting strength $\kappa$ the barrier height
can be estimated by assuming that only one unit of the chain is
elongated, which yields a value
$\Delta E_{\kappa}=(\omega_0^2+2\kappa)^3/(6a^2)$.
Hence, compared to
the isolated unit, a unit coupled to its neighbors experiences an
increase of the barrier height.

For sufficiently small energy per unit of all chain members as
compared to the potential barrier, a linear regime with $a=0$ holds true in the
considered potential (\ref{eq:potential}), yielding oscillatory
solutions in phase space such that the elongations are restricted
to the neighborhood of the potential bottom. The corresponding linearized system
\begin{equation}
\delta\ddot{q}_n+\omega_0^2{\delta{q}}_n-\kappa\,\left[\,{\delta{q}}_{n+1}+{\delta{q}}_{n-1}-2{\delta{q}}_n\right]=0
\,,\label{eq:linear}
\end{equation}
possesses exact plane-wave solutions (phonons)
\begin{equation}
{\delta{q}}_n(t)={{u}_0} \exp[i(k\,n-\omega\,t)]+c.c.
\,,\label{eq:solution}
\end{equation}
%
%
%
%
The wave number $k=2\pi m/N$, with integer $m\in [-N/2,N/2]$, and
the frequency $\omega$ are related by  the dispersion relation
\begin{equation}
\omega^2=\omega_0^2+4\kappa
\,\sin^2\left(\frac{k}{2}\right)\,.
\end{equation}
This expression determines the frequency of linear oscillations in
the phonon band with $k \in [-\pi, \pi]$.

The superposition of phonon modes causes oscillatory states wherein
distinguished units may temporarily accumulate energies that are
comparable to the barrier energy. However, in a harmonic potential
with $a=0$ these states are highly unlikely. If at all, they occur
at a time scale comparable to the Poincar\'{e} recurrence time of
the system.

Nonetheless, utilizing nonlinear effects with $a > 0$, an initial
state near the metastable minimum is structurally unstable which
mobilizes structural transition of the chain such that a transition
state is adopted. This mechanism will be elaborated on in the next
section.

\section{Modulational instability}\label{section:modulational}

It is well established that the formation of localized excitations
in nonlinear systems can be caused by a modulational instability
\cite{Kivshar92}-\cite{Peyrard98}. This mechanism initiates an
instability of an initial plane-wave when small perturbations of
non-vanishing  wave numbers are imposed. The instability, giving rise
to an exponential growth of the perturbations, destroys the initial
wave at a critical wave number, so that a localized hump is formed.

To analyze the nonlinear character of the solutions of
Eq.\,(\ref{eq:qdot}) a nonlinear discrete Schr\"{o}dinger equation
for the slowly varying envelope solution, $u_n(t)$, has been derived
in \cite{Kivshar93,Daumont}
\begin{equation}
2i \omega_0\,\dot{u}_n+\kappa\,\left[u_{n+1}+u_{n-1}-2u_n\right]
+\gamma\,|u_n|^2 u_n=0\,,\label{eq:DNLS}
\end{equation}
with the nonlinearity parameter
$\gamma={10a^2}/{3\omega_0^2}$. The stability of a plane-wave solution of
Eq.\,(\ref{eq:DNLS}) of the form 
\begin{equation}
u_n(t)=u_0\,\exp(i\theta_n)+c.c.\,,\label{eq:uq0}
\end{equation}
with $\theta_n=k n-\omega t$ can be investigated in the weakly
nonlinear regime by assuming small perturbations of the amplitude
$u_0$ and phase $\theta_n$ that have the form of sinusoidal
modulations with wave number $Q$ and frequency $\Omega$.
One then
finds  for the perturbational wave the following dispersion relation
\cite{Kivshar93},\cite{Daumont}:
\begin{eqnarray}
\lefteqn{\left[\Omega-2\kappa\,\sin(Q)\sin(k)\right]^2=}\nonumber\\
&=&\frac{2\kappa}{\omega_0^2}\,\sin^2\left(\frac{Q}{2}\right)\cos(k)\,\nonumber\\
&\times&\left[2\kappa\,\sin^2\left(\frac{Q}{2}\right)\cos(k)-\gamma
u_0^2\right]\,.\label{eq:modulational}
\end{eqnarray}
Stability of the perturbations necessitates that $\Omega$ is real.
Conversely, if the right hand side in Eq.\,(\ref{eq:modulational}) is
negative the perturbation grows exponentially with a rate
\begin{equation}
\Gamma=\left[\frac{2\kappa}{\omega_0^2}\,\sin^2\left(\frac{Q}{2}\right)\cos(k)
\left(\gamma\,u_0^2-2\kappa\,\sin^2\left(\frac{Q}{2}\right)
\cos(k)\right)\right]^{1/2}\,.
\label{eq:growthrate}
\end{equation}

Notably, this modulational instability is  possible only  in the
range of carrier wave numbers $k\in [0,\pi/2)$. Thus, patterns of
short wave length are insensitive with respect to modulations.

\begin{figure}
  \begin{center}
    \begin{minipage}{8.cm}
      \resizebox{8.cm}{6.cm}{\includegraphics[scale=1.0]{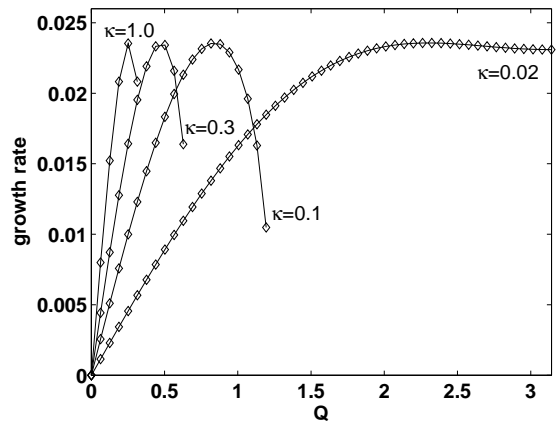}}
    \end{minipage}
    \begin{minipage}{5.cm}
    \end{minipage}
  \end{center}
\caption{\label{fig:growth}
 Growth rate in
dependence of the wave number $Q$ for different coupling strengths
$\kappa$, as indicated on the graphs. The parameter values are
$k=0$, $u_0=0.2$, $a=1$, $\omega_0^2=2$, and $N=100$.}
\end{figure}

In the following we focus our interest on the $k=0$ mode. In
Fig.~\ref{fig:growth}
we depict the growth rate for
different values of the coupling strength for a mode with $u_0=0.2$, $k=0$
and $N=100$. The inequality
\begin{equation}
\gamma\,u_0^2-2\kappa\,\sin^2\left(\frac{Q}{2}\right)\ge 0
\label{eq:constraint_Q}
\end{equation}
puts a constraint on the allowed wave numbers. For relatively small
coupling strength  $\kappa=0.02$ the whole range of wave numbers
$|Q|\leqslant \pi$ is responsible for the modulational instability,
albeit with different weights. Enlarging $\kappa$ not only
increasingly shifts the cut-off for the allowed wave numbers towards
$Q=0$ but in addition makes the instability bands also narrower. In other
words the modulational instability becomes more mode-selective.
Nevertheless, the maximum of the growth rate
\begin{equation}
\Gamma_{max}=\frac{1}{2\omega_0}\gamma u_0^2\,,\label{eq:rate}
\end{equation}
is unaffected by alterations of $\kappa$, while its position
\begin{equation}
Q_{max}=2\sin^{-1}\left(\sqrt{\frac{\gamma
u_0^2}{4\kappa}}\right)\,,
\end{equation}
moves closer to zero with increasing coupling strength   $\kappa$.

The way the growth rates with corresponding weights for
perturbations at different wave numbers $Q$ contribute to a mean
growth rate is determined by the  quantity $\bar\Gamma$, reading:
\begin{equation}
\bar{\Gamma}=\frac{2}{N}\sum\limits_{n=0}^{N/2}\Gamma
\left(Q=\frac{2\pi}{N}\,n,k=0\right)\,.
\end{equation}
\begin{figure}
  \begin{center}
    \begin{minipage}{8.cm}
      \resizebox{8.cm}{6.cm}{\includegraphics[scale=1.0]{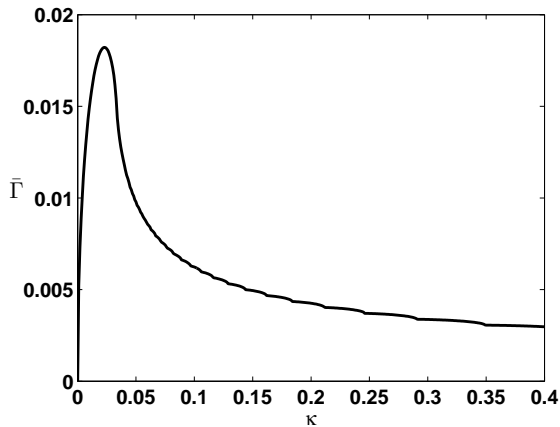}}
    \end{minipage}
    \begin{minipage}{5.cm}
    \end{minipage}
  \end{center}
\caption{\label{fig:meangrowth} Mean growth rate versus the coupling
strength. The parameter
values used are the same as in Fig.~\ref{fig:growth}.}
\end{figure}
This quantifier is depicted in Fig.~\ref{fig:meangrowth} as a
function of the coupling strength $\kappa$. The maximum around
$\kappa
\simeq 0.023$
suggests that a sizable divergence of the perturbations is induced.
To the left of its maximum the mean growth rate drops drastically
while on its right the decrease considerably weakens.

\section{Energy sharing and formation of arrays of
breathers}\label{section:breather}

In order to enhance the energy localization in the dynamics of
Eq.\,(\ref{eq:qdot}) we propose the following scenario: An  amount
of energy $ E_0 = E_{total}/N $ is applied per unit  which allows
the activation of nonlinear, self-organized excitations of the
chain.

Thus, the chain possesses a total  energy $E_{total}=N E_0$. For an
escape to take place we  have:  $E_{total}> E_{act}> \Delta E$. This
inequality conveys the fact that more than just one unit governs the
escape mechanism. The initial energy $E_0$ is supplied as follows:
(i) First, the whole chain is elongated homogeneously along a fixed
position  $q_0$ near the bottom of the well. 
[Notice that from Eq.\,(\ref{eq:uq0}) it follows that 
this corresponds to a flat mode being equivalent to a $k=0$ plane-wave solution and thus $q_0=2u_0$.]
(ii) Then,  the
position of all units and  their momenta are iso-energetically
randomized while  keeping the total energy a constant---i.e.
$E_{total}=N E_0=\mbox{const} $. 

The random position values are
chosen from a bounded interval $|q_n(0)-{q_0}| \leq \Delta q$ and,
likewise, the random initial momenta, $|p_n(0)-p_0|\leq \Delta p$. From Eq.\,(\ref{eq:solution}) for a plane-wave solution with wave number $k=0$ 
one deduces that $p_0=0$.
The whole chain is thus initialized close to an {\it almost
homogeneous state}, but yet sufficiently displaced  ($\Delta q \ne
0$) in order to generate nonvanishing interactions, enabling the
exchange of energy among the coupled units.

\begin{figure}
  \begin{center}
    \begin{minipage}{8.cm}
      \resizebox{8.cm}{7.cm}{\includegraphics[scale=3.]{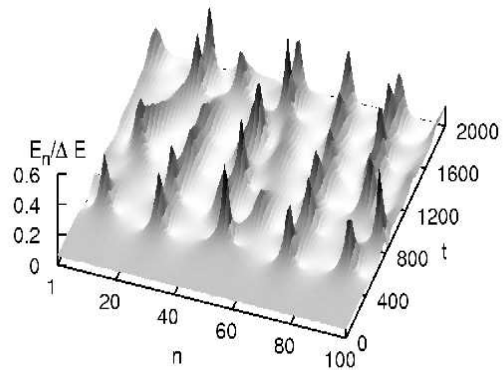}}
    \end{minipage}
    \begin{minipage}{5.cm}
    \end{minipage}
  \end{center}
\caption{\label{fig:density} Spatio-temporal evolution of the energy
distribution $E_n(t)$. Initially, the coordinates are uniformly
distributed $(k=0)$ within the interval $|q_n(0)-{q_0}|\le \Delta q$
with an average at ${q_0}=0.3$ and width $\Delta q=0.01$ and zero 
momenta---i.e., $p_0=\Delta p=0$. This yields a total energy
$E_{total} = 8.1\equiv 6.075\Delta E$. The parameter values
are given by $a=1$, $\omega_0^2=2$, $N=100$, and
 $\kappa=0.3$.}
\end{figure}

As the role of the deviation of the initial conditions from a
completely homogeneous state for the instigation of the energy
exchange process is concerned
%
we observe that
the attainment of an array of humps  proceeds
the faster the
larger
 is the width $\Delta p$ and/or
$\Delta q$.
More precisely, due to the emergence of a modulational instability a
pattern evolves in the course of time (of the order of $t\sim
5\times 10^2$) where for some lattice sites the amplitudes grow
considerably whereas they remain relatively small in the adjacent
regions. This feature is illustrated in Fig.~\ref{fig:density},
depicting the spatio-temporal evolution of the site-energy:
\begin{equation}E_n=\frac{p_{n}^{2}}{2}
\,+\,U(q_n)\,+\,\frac{\kappa}{4}
\left[\,\left(\,q_{n+1}-q_{n}\,\right)^2+\left(\,q_{n-1}-q_{n}\,\right)^2\,\right]
\label{eq:endensity}\,.
\end{equation}
The breather states possessing a relatively high energy occur
spontaneously at an average distance of the inverse wave numbers
$Q_{max}^{-1}$, corresponding to the maximal growth rate
$\Gamma_{max}$ in (\ref{eq:rate}). Upon moving, these breathers tend
to collide inelastically with others. In fact, various breathers
merge to form larger-amplitude breathers, proceeding preferably such
that the larger-amplitude breathers grow at the expense of the
smaller ones. As a result, a certain amount of the total energy
becomes strongly concentrated within confined regions of the chain.
This localization scenario is characteristic for nonlinear lattice
systems \cite{Daumont}, \cite{Bang}-\cite{Tsironis}.

For our simulations the set of coupled equations (\ref{eq:qdot}) has
been numerically integrated by use of a fourth-order Runge-Kutta
scheme. The accuracy of the calculation was checked by monitoring
the conservation of the total energy with precision of at least
$10^{-5}$. The investigated chain consists of $100$ coupled
oscillators.

To relate the energy localization with an escape over the barrier we
note that in the beginning the
average amount of energy contained in a single unit,
$E_0={E_{total}}/{N}$, lies significantly below the barrier energy
as expressed by the low ratio $E_0/\Delta E=0.06$. Thus, a single
unit must acquire  the  energy content of at least $16$ nearby units
before it is able to overcome the barrier.

\begin{figure}
  \includegraphics[scale=0.5]{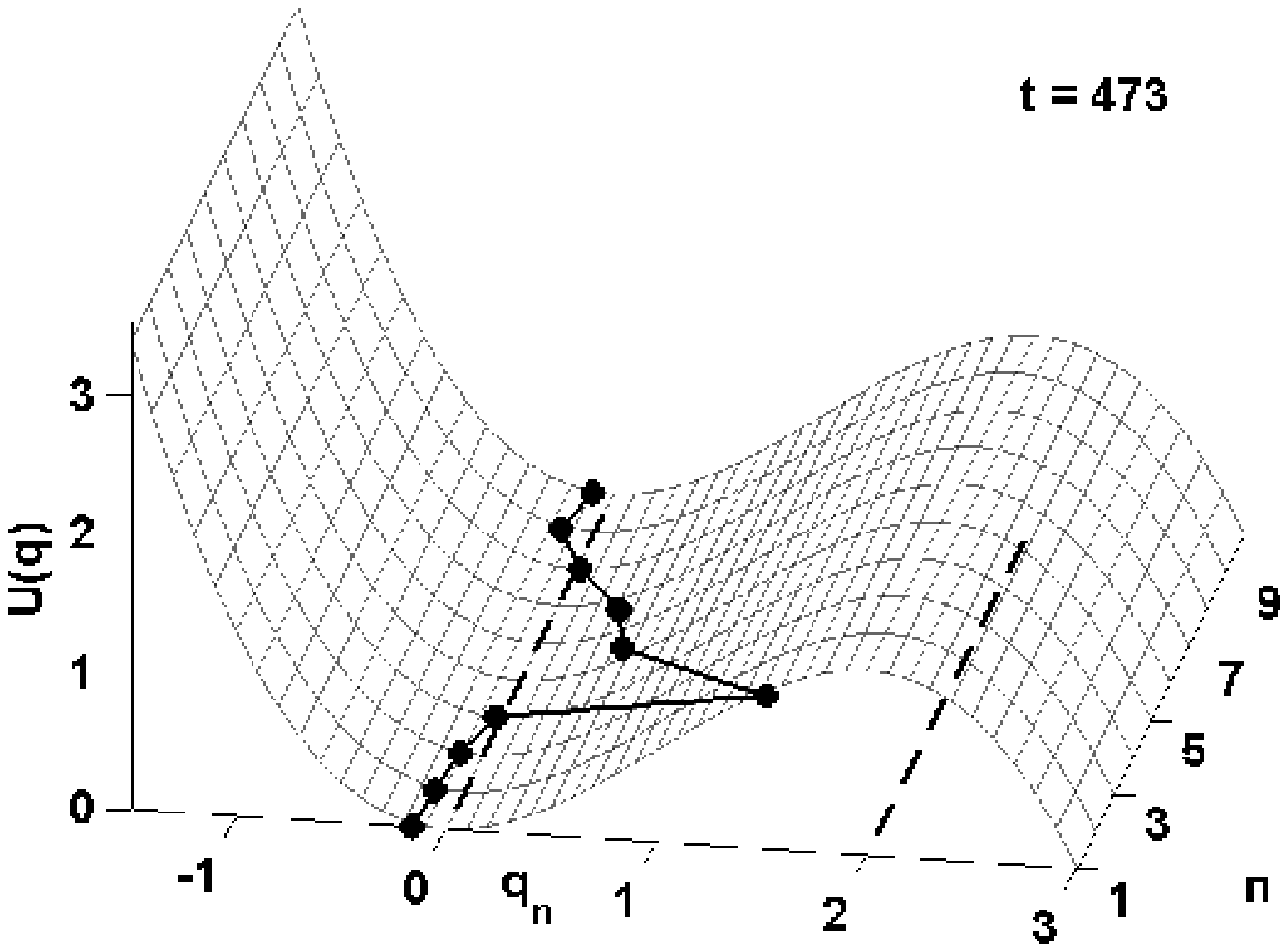}
  \includegraphics[scale=0.5]{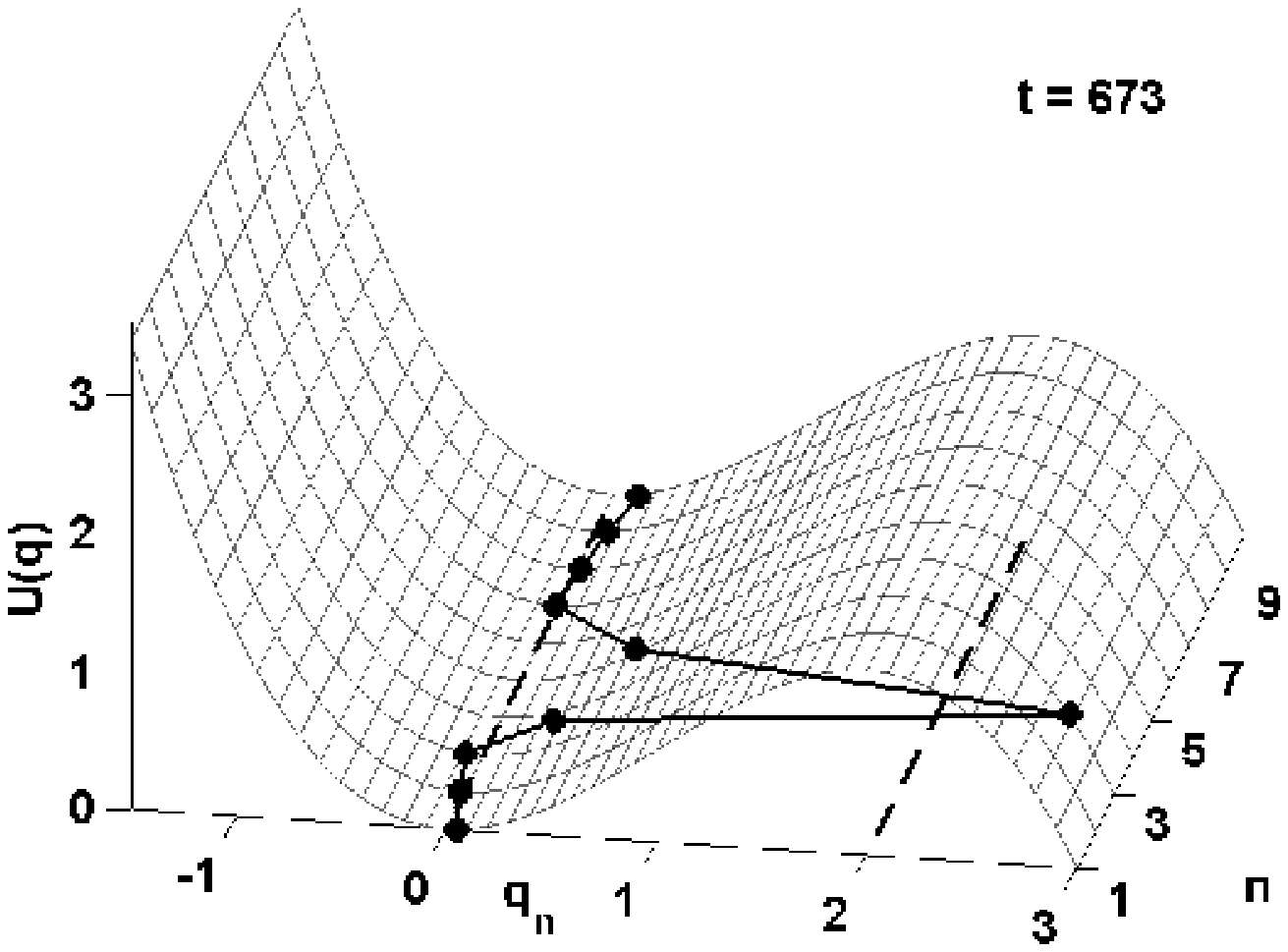}
  \caption{\label{fig:snap} Snapshots of the amplitudes $q_n(t)$ of a segment of the lattice
  covering ten oscillators, illustrating the formation of a localized structure
    with a subsequent barrier crossing event of the chain.  Initial
    conditions are $q_0=0.45$, $\Delta q=0.01$, and $p_0=\Delta p=0$.
    Parameter values are $\omega_0^2=2$, $a=1$, $N=100$, and $\kappa=0.2$.
    Top panel: snapshot taken at time $t=473$: Emergence of localized structures.
Bottom panel: snapshot taken at time $t=673$: Depicted is a segment
of the lattice with one unit located beyond the barrier. The dashed
lines in the ($q_n-n$)-plane designate the position of the
potential's minimum and maximum.}
\end{figure}

For further illustration we depict in Fig.~\ref{fig:snap}
snapshots of $q_n(t)$ at two different instants of time.
In the beginning the energy is virtually equally shared among all
units (not shown). After a certain time has evolved, the local
energy accumulation is  enhanced in such a manner that at least one
of the involved units possess enough energy to overcome the barrier.
The question then is, does such an escaped unit continue its
excursion beyond the barrier or can it even be pulled back into the
bound chain formation ($q_n< q_{max}$) by the restoring binding
forces exercised by its neighbors? On the other hand, the unit that
has already escaped from the potential well might drag neighboring
ones closer to or, in the extreme, even over the barrier. Thus, a
concerted escape of the chain
from the potential valley becomes plausible .


\section{Transition state}\label{section:transition}

\begin{figure}
  \begin{center}
    \begin{minipage}{8.cm}
      \resizebox{8.cm}{6.cm}{\includegraphics[scale=1.0]{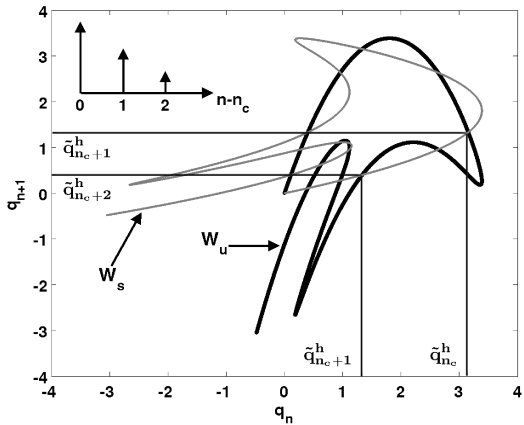}}
    \end{minipage}
    \begin{minipage}{5.cm}
    \end{minipage}
  \end{center}
\caption{\label{fig:manifold} First intersections of
the stable $W_s$ (thin gray line) and the unstable
$W_u$  manifold   (thick dark line) of the hyperbolic
point of the map in Eq. (\ref{map}) at the map origin $(0,0)$. The
parameter values are $\omega_0^2=2$, $a=1$, and $\kappa=1$. The
homoclinic point labeled $(\tilde{q}_{n_c}^h;\tilde{q}_{n_c+1}^h)$ and its map iterate $(\tilde{q}_{n_c+1}^h;\tilde{q}_{n_c+2}^h)$
correspond on the consecutive lattice sites $n=n_c,n_c+1,n_c+2$ to a
decaying pattern whose amplitudes with $\tilde{q}_{n}^h>\tilde{q}_{n+1}^h$ are
represented by the arrows of varying length in the schematic
representation
 in the upper left corner.
Map iterations of the homoclinic point $(\tilde{q}_{n_c+1}^h;\tilde{q}_{n_c+2}^h)$ result in further homoclinic points approaching asymptotically
the hyperbolic point at the map origin.}
\end{figure}

\begin{figure}
  \begin{center}
    \begin{minipage}{8.cm}
      \resizebox{8.cm}{6.cm}{\includegraphics[scale=1.0]{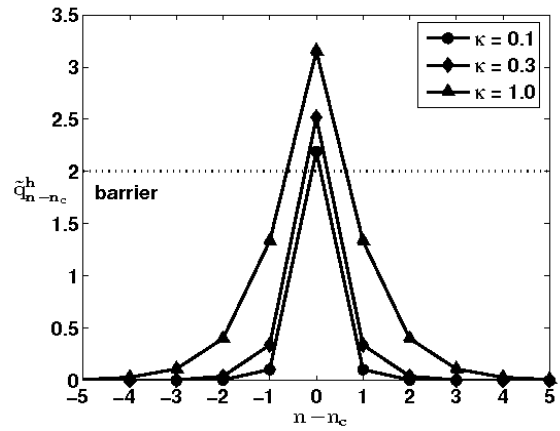}}
    \end{minipage}
    \begin{minipage}{5.cm}
    \end{minipage}
  \end{center}
\caption{\label{fig:profile} Amplitude profile of the critical chain
configuration, being symmetrically centered at site $n_c$, for different coupling strengths $\kappa=0.1$
(circles), $\kappa=0.3$ (diamonds), and $\kappa=1.0$ (triangles). For
better illustration only  we depict the central part of the lattice
chain with sizable elongations. The parameter values are
$\omega_0^2=2$ and $a=1$.}
\end{figure}

Whether a unit of growing amplitude can in fact escape from the
potential well or is held back by the restoring forces of their
neighbors depends on the corresponding amplitude ratio as well as on
the coupling strength. The critical chain configuration---that is, the
{\it transition state} separating bounded ($q_n<q_{max}$) from
unbounded ($q_n>q_{max}$) lattice solutions---is determined by
$\ddot{q}_n(t)=0$. The system of equations (\ref{eq:qdot}) then  reduce to the
stationary system of equations:
\begin{equation}
-\frac{\partial U}{\partial \tilde{q}_n}+\kappa[\tilde{q}_{n+1}+\tilde{q}_{n-1}-2\tilde{q}_n]=0\,.\label{eq:difference}
\end{equation}
Interpreting $n$ as a 'discrete' time, with $1\le n\le N$, equation
(\ref{eq:difference})
describes the motion of a point particle in
the inverted potential $-U(\tilde{q})$. This difference system can
be cast in the form of a two-dimensional map by defining $x_n=\tilde{q}_n$ and
$y_n=\tilde{q}_{n-1}$ \cite{Physicsreports}, which gives
\begin{eqnarray}
x_{n+1}&=&(\omega_0^2x_n-ax_n^2)/\kappa+2x_n-y_n\nonumber\\
y_{n+1}&=&x_n\,. \label{map}
\end{eqnarray}
The fixed points of this map are found as
\begin{equation}
x_0=y_0=0\,,\qquad \qquad x_1=y_1=\frac{\omega_0^2}{a}\,.
\end{equation}
A linear stability analysis reveals that $(x_0,y_0)$ represents an
unstable hyperbolic equilibrium while at $(x_1,y_1)$ a stable
center is located.
The map is non-integrable. The stable and unstable manifolds of the
hyperbolic point intersect each other, yielding homoclinic crossings
as illustrated in Fig.~\ref{fig:manifold}.
The corresponding
homoclinic orbit of the map,
consisting of the points of intersections between $W_s$ and $W_u$, is
on the lattice chain equivalent to a stationary localized hump
solution $\{\tilde{q}_{n}^{h}\}$, centered at site $n=n_c$, which resembles the form of a (pointed)
 hairpin (for details concerning the relation between homoclinic orbits
and localized lattice solutions see \cite{Physicsreports} and
\cite{PRE96}).
In Fig.~\ref{fig:profile} profiles of this hairpin-like {\it critical localized mode} (c.l.m.), or critical
nucleus, with displacements
$\{\tilde{q}_{n}^{h}\}$
are depicted for several coupling strengths.
We observe that the
stronger the coupling is, the larger the maximal amplitude of the humps is, $\tilde{q}_{n_c}^h=\tilde{q}_{max}^h$, and the wider the spatial extension of the latter is. We
underline that on a sufficiently extended lattice this c.l.m. represents a narrow chain formation with its width being much
smaller than the total chain length.
Equation (\ref{eq:difference}) can be derived from an energy
functional
with vanishing kinetic energy,
reading
\begin{equation}
E[\{\tilde{q}_n\}]=\sum_n\,\left(U(\tilde{q}_n)+\frac{\kappa}{2}[\tilde{q}_n-\tilde{q}_{n-1}]^2\right)
\end{equation}
with $\partial E/\partial \tilde{q}_n=0$. Apparently, with
increasing coupling strength $\kappa$ a larger activation energy
\begin{equation}
E_{act}=E[\{\tilde{q}_n^h\}]
\end{equation}
is required to bring the chain into its critical localized mode
configuration.

\begin{figure}
  \begin{center}
    \begin{minipage}{8.cm}
      \resizebox{8.cm}{6.cm}{\includegraphics[scale=1.0]{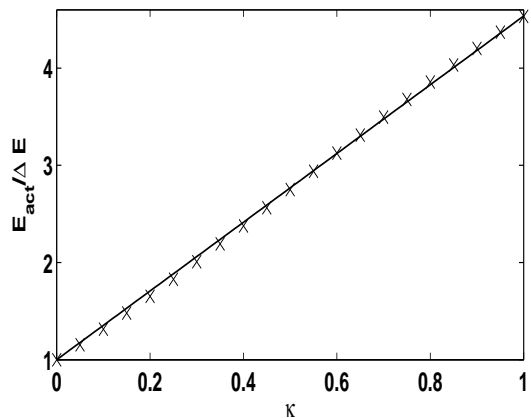}}
    \end{minipage}
    \begin{minipage}{5.cm}
    \end{minipage}
\end{center}
\caption{\label{fig:ae}Activation energy $E_{act}$ as a function of the coupling strength $\kappa$.
 Solid line: fitted data with slope $s=3.54\pm 0.01$. The remaining parameter values are $\omega_0^2=2$ and $a=1$.}
\end{figure}

The activation energies for different coupling strengths are depicted in Fig.~\ref{fig:ae}.
For small coupling the maximal amplitude $\tilde{q}_{max}^h$
is beyond but still
close to $q_{max}=\omega_0^2/a$
[the position of the maximum of the potential $U(q)$] while the
remaining units practically reside  at the minimum of the potential
$U(q)$. For larger couplings the maximal amplitude
$\tilde{q}_{max}^h$ lies at larger distances beyond $q_{max}$. Most
importantly, for escape we have that $\partial U(\tilde{q})/\partial
\tilde{q}|_{\tilde{q}=\tilde{q}_{max}^h}<0$.

\begin{figure}
  \begin{center}
    \begin{minipage}{8.cm}
\resizebox{8.cm}{6.cm}{\includegraphics[scale=1.]{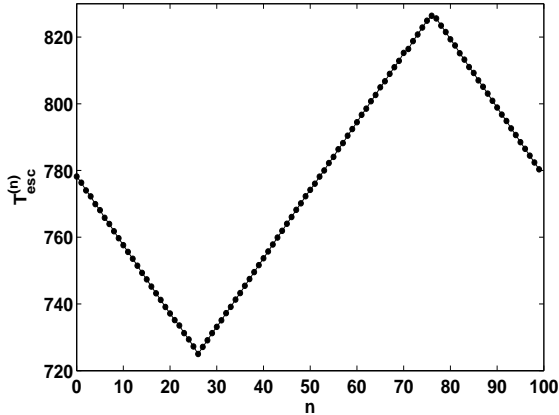}}
    \end{minipage}
    \begin{minipage}{5.cm}
    \end{minipage}
  \end{center}
\caption{\label{fig:timecoll} The escape time $T_{esc}^{(n)}$ of the
individual chain units versus position for collective escape and
periodic boundary conditions for a chain consisting of $100$ units for one realization of initial conditions.
The parameter values are $a=1$, $\omega_0^2=2$, $N=100$, and
$\kappa=0.3$. The initial conditions are given by $q_0=0.45$,
$\Delta q=0.1$, and $p_0=\Delta p=0$.}
\end{figure}

It can be shown that the critical localized mode,
being associated with an unstable saddle point in configuration space, is indeed dynamically unstable.
Setting $q_n(t)=\tilde{q}_n^h+w_n(t)$ with $|w_n|\ll 1$  the linearized equations of motion are derived as
\begin{eqnarray}
\ddot{w}_n(t)&=&-\frac{\partial^2 U({q}_n)}{\partial q^2_n} \rvert_{q_n={\tilde q}_n^h} w_n(t)\nonumber\\
&+&\kappa [w_{n+1}(t)+w_{n-1}(t)-2 w_n(t)]
\,.\label{eq:linearized}
\end{eqnarray}
With the ansatz $w_n(t)=\phi_n\exp(\sqrt{\lambda} t)$ for the solution of (\ref{eq:linearized})
 one arrives at an eigenvalue problem
\begin{equation}
\lambda \phi_n=-V_n\phi_n+\kappa[\phi_{n+1}+\phi_{n-1}-2\phi_n]\,,\label{eq:tight-binding}
\end{equation}
with
\begin{equation}
V_n=\frac{\partial^2U({q}_n)}{\partial q^2_n}\rvert_{q_n={\tilde q}_n^h}=\omega_0^2-2a\tilde{q}_n^h\,.
\end{equation}
The second-order difference equation (\ref{eq:tight-binding})
is of the discrete stationary Schr\"{o}dinger type,
 with a
non-periodic potential $-V_n$, breaking the translational
invariance so that localized solutions exist (so called stop-gap
states). The evolution of the two-component vector
$(\phi_{n+1},\phi_n)^T$ is determined by the following Poincar\'{e}
map:
\begin{eqnarray}
\cal{M}:& &
\begin{pmatrix}
\phi_{n+1} \\
\phi_n
\end{pmatrix}=
\begin{bmatrix}
E_n & -1\\1 & 0
\end{bmatrix}
\begin{pmatrix}
\phi_{n} \\
\phi_{n-1}
\end{pmatrix}\,,
\end{eqnarray}
with on-site energy $E_n=(\lambda+V_n)/\kappa+2$.
The nodeless even-parity
ground state of Eq.\,(\ref{eq:tight-binding}), with its energy
under the lower edge of the energy band of the passing band states,
corresponds to an orbit of the linear map $\cal{M}$ being
homoclinic to the hyperbolic equilibrium point
at the origin $(0,0)$ of the map plane.
For the presence of a  hyperbolic equilibrium the following inequality has to be satisfied:
\begin{equation}
\Trace({\cal{M}})=E_n=\frac{\lambda+V_n}{\kappa}+2>2\,,
\end{equation}
implying that $\lambda$ must fulfill the constraint
\begin{equation}
\lambda > \max_n (-V_n)=2a \max_n \tilde{q}_n^h-\omega_0^2\,>0.
\end{equation}
With the maximal amplitude of the c.l.m. lying beyond the barrier---viz., $\max_n \tilde{q}_n^h>\omega_0^2/a$---one finds
\begin{equation}
\lambda >\omega_0^2>0\,.
\end{equation}
Therefore, the ground state belongs to a positive eigenvalue
from which we deduce that perturbations of the corresponding solution in the time domain grow exponentially.
Hence, if the kinetic energy overcomes the critical nucleus, the
subsequent escape of its neighbors
is
initiated, which progress on the chain to the left and to the right
of the hair pin as a propagating kink and anti-kink, respectively
(see Refs. \cite{Sebastian,LanBut,MarHan1,MarHan2}). In phase space
the units move parallel to the unstable manifold of the hyperbolic
equilibrium [which is related to the saddle point at the maximum of
the potential $U(q)$], realizing in this way an efficient lowering
of the total potential
 energy.  Because the kinetic energy of this outward motion is
consequently increasing, a return backwards over the  barrier into
the original well is hereby prevented. Fig.~\ref{fig:timecoll}
illustrates the kink-antikink motion
showing the escape time $T_{esc}^{(n)}$ of the units versus the position
on the lattice. The {\it escape time of a unit is defined as the
moment at which it passes through the value $q_{esc}=5\,q_{max}$
beyond the barrier}.
We remark that $q_{esc}$ is chosen such that $U(q_{esc})$ is
sufficiently lowered so that the return of an escaped unit over the
barrier into the potential well is practically excluded.
Consecutively,
all oscillators manage to eventually climb over the barrier one
after another in a relatively short time interval. (The
position of the first escape event varies in general  for random samples of initial conditions.)

\section{Comparison with thermally activated escape}\label{section:comparison}

We next compare the microcanonical escape process with the
corresponding thermally assisted escape process at a temperature $T$
\cite{RMP,JSPHa,Sebastian,Lee,Kraikivsky,LanBut,MarHan1,MarHan2}.
The associated Langevin system reads
\begin{equation}
\frac{d^2q_{n}}{dt^2}\,+\,\frac{dU}{dq_n}-\kappa\,\left[\,q_{n+1}+q_{n-1}-2q_n\,\right]+\nu\frac{d q_n}{dt}+\xi_n(t)
=0\,\,,\label{eq:canonical}
\end{equation}
with a friction parameter $\nu$ and  where $\xi_n(t)$ denotes a
Gaussian distributed thermal, white noise of vanishing mean
$<\xi_n(t)>=0$, obeying the well-known  fluctuation-dissipation
relation $<\xi_n(t) \xi_{n^{\prime}}(t^{\prime})>=2\nu k_B
T\delta_{n,n^{\prime}}\delta(t-t^{\prime})$ with $k_B$ denoting the
Boltzmann constant.
We define the {\it escape time of a chain} as the mean value of the
escape times of its units (see again above).

\subsection*{Escape times}
\begin{figure}
  \begin{center}
    \begin{minipage}{8.5cm}
\resizebox{8.5cm}{6.cm}{\includegraphics[scale=1.]{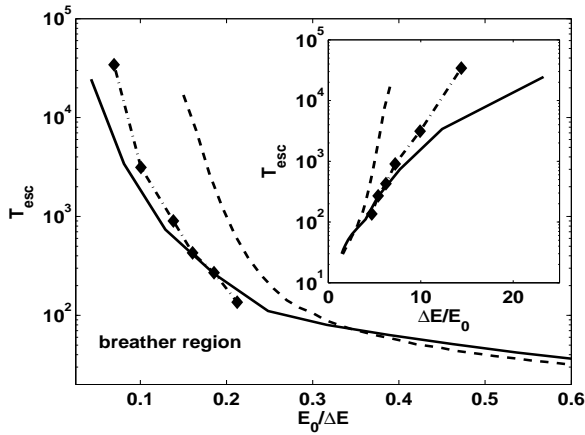}}
    \end{minipage}
    \begin{minipage}{5.cm}
    \end{minipage}
  \end{center}
  %
  %
\caption{\label{fig:meantime} Mean escape
      time $T_{esc}$ {\it vs.} the ratio $E_{0}/\Delta E$ of energy per unit $E_0$ and the barrier energy $\Delta E$
       for the noise-free case with initial $k=0$ mode (solid line) with
       $\Delta q=0.01$ and $p_0=\Delta p =0$ (for $q_0$, see text), the case with strongly randomized initial conditions
       (diamonds) where $-0.251\le q_n(0) \le 0.534$, $-0.251\le p_n(0) \le 0.534$,
      and the noise-assisted escape  (dashed line), respectively.
      The parameter values are $\omega_0^2=2$, $a=1$, $\kappa=0.3$,
      and the friction strength is
$\nu=0.1$. The inset depicts the same data, but now plotted versus
the effective Arrhenius factor $\Delta E/E_0$.}
\end{figure}

Our results are summarized in Fig.~\ref{fig:meantime}
depicting the mean escape time as a function of $E_{0}/\Delta E$.
The  averages were performed over $500$ realizations of random
initial conditions or the noise in the microcanonical  and Langevin
systems, respectively. For the deterministic and conservative system
(\ref{eq:qdot}) the excitation energy $E_0$ is given by the
(average) initial energy content of one unit $E_0=E_{total}/N$.
For the simulations we varied $q_0$ while keeping $\Delta q=0.01$ and $p_0=\Delta p=0$ and obtained thus different values of
$E_{total}$. In the case of
the Langevin system (\ref{eq:canonical}) for sufficiently low $T$ the energy $E_0$
is taken as  the thermal energy $k_BT$. This ratio thus corresponds
to the inverse Arrhenius factor \cite{RMP}; indeed, at sufficient low
temperature (large ratio $\Delta E/E_0$) the logarithmic escape time
follows an almost linear behavior versus the Arrhenius factor, as
expected for a noise-driven escape at weak noise strength
\cite{RMP}.

The Langevin equations were numerically integrated using a two-order
Heun stochastic solver scheme \cite{Langevin_Numerics}. In both
cases there occurs a rather distinct decay of $T_{esc}$ with growing
ratio $E_0/\Delta E$ in the low energy region. This effect weakens
gradually upon further increasing $E_0$. Remarkably, for low $E_0$
(indicated as the breather region in the plot) the escape proceeds
 distinctly faster for the noise-free case as compared with a situation of a
chain that is coupled to a heat bath at temperature $T$. This
implies a large enhancement of the rate of escape as compared to the
thermal rate. Near $E_0/\Delta E\ge 0.36$, there occurs a crossover,
with the mean escape time of the deterministic system at even higher
ratios closely following that of the thermal Langevin dynamics.
At these values the escape times become comparable with the relaxation time which is determined by
the inverse of the friction strength. Apparently in the region of lower $k_BT$
 nonlinear excitations are damped out in the Langevin system at longer time scales. Hence they will not
accelerate the escapes in the case with fluctuations and damping.


To sharpen our finding  that the escape proceeds typically faster in
the noiseless situation as compared to the case with a coupling to a
heat bath, we investigated also the escape process of non-flat chain patterns
starting out from {\it strongly randomized} initial conditions.
For these initial conditions the coordinates and
momenta are chosen at random from
fairly broad ranges $-0.251\le q_n(0) \le 0.534$ and $-0.251\le p_n(0) \le 0.534$.
For various energy values
 the averages were performed over $100$ realizations
of initial conditions  belonging to iso-energetic configurations
with ratio $E_0/\Delta E$ each.

The findings for the mean
escape time as a function of  the mean initial energy content of the
units relative to the barrier height $E_0/\Delta E$ are included in
Fig.~\ref{fig:meantime} with the diamond symbols.
Most importantly, even for random initial conditions the mean escape
time assumes smaller values  in the microcanonical situation as
compared to the Langevin dynamics. This underpins  our general
statement that noiseless escape indeed proceeds faster than
thermally activated escape.

We note that the breathers present robust chain configurations that
are formed rather fast as compared to the escape time. In contrast,
the forever impinging stochastic forces seemingly impede  a fast
growth of the critical nucleus and may even cause a possible
destruction of the critical chain formation, leading to re-crossings
of the transition region, which only hampers a speedy escape. This
inhibition for escape is most effective at small ratios of
$E_0/\Delta E$, being induced either by high barrier heights, or low
temperatures (implying a small $E_0$). A deterministic scenario thus
presents a more favorable route towards accelerated escape in
situations with very weak noise or very large barrier heights.
Having performed also simulations for more general situations (i) with 
nonharmonic, nonlinear chain interactions, (ii) in higher
dimensions, and (iii) with differing on-site potentials, we find
\cite{preparation} that the phenomenon of an enhanced, noise-free
escape remains robust in regimes of a large effective Arrhenius
factor with the latter given by the ratio of the barrier height
$\Delta E$ and the initial energy per unit, $E_0$.


\section{Optimal coupling and resonance structure in the escape process}\label{section:resonance}

\begin{figure}
  \begin{center}
    \begin{minipage}{8.cm}
      \resizebox{8.cm}{6.cm}{\includegraphics[scale=1.0]{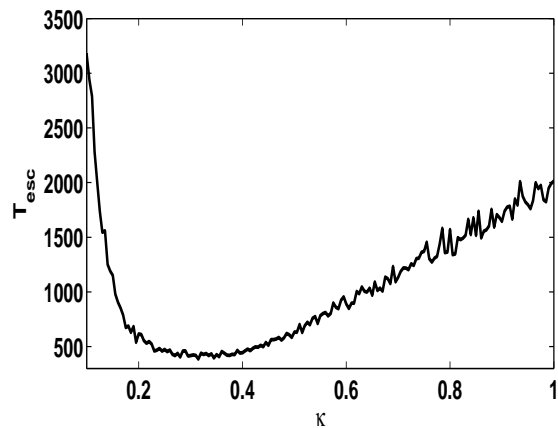}}
    \end{minipage}
  \end{center}
\caption{\label{fig:resonance} Resonant behavior in the
mean
escape time $T_{esc}$ as a function of the coupling strength
$\kappa$ with the remaining parameter values as in
Fig.~\ref{fig:density}. The initial conditions are for the
coordinates $q_0=0.45$ and $\Delta q=0.1$ and for the momenta
$p_0=\Delta p=0$.
The  averages were performed over $500$ realizations of random
initial conditions.}
\end{figure}

Furthermore, we study the impact of the coupling strength $\kappa$ on
the escape process. The results concerning the mean escape time are
illustrated with Fig.~\ref{fig:resonance}.
Strikingly, the mean escape time exhibits a {\it resonance
structure}; viz., there exists a coupling strength
($\kappa_{res}=0.31$) for which the escape proceeds faster than for
all other couplings strengths. Upon lowering $\kappa <\kappa_{res}$
we notice a substantial rise of the escape time while for $\kappa >
\kappa_{res}$ the graph exhibits only a moderately growing slope
with growing coupling strength $\kappa$. In this sense
$\kappa_{res}=0.31$ from Fig.~\ref{fig:resonance}\, represents
indeed the {\it optimal} coupling strength for which  escape is
achieved within a minimal amount of time. Finally, outside the range
$\kappa \in [0.05\,,\,1.5]$ not even the escape of a single unit has
been observed. The reason is that the time scale for a pronounced
formation of energy concentration, being vital for escape (due to
breather coalescence and energy accumulation in the critical
localized mode), exceeds the simulation time (taken here as $t=5000$).

\begin{figure}
  \begin{center}
    \begin{minipage}{8.cm}
      \resizebox{8.cm}{6.cm}{\includegraphics[scale=1.0]{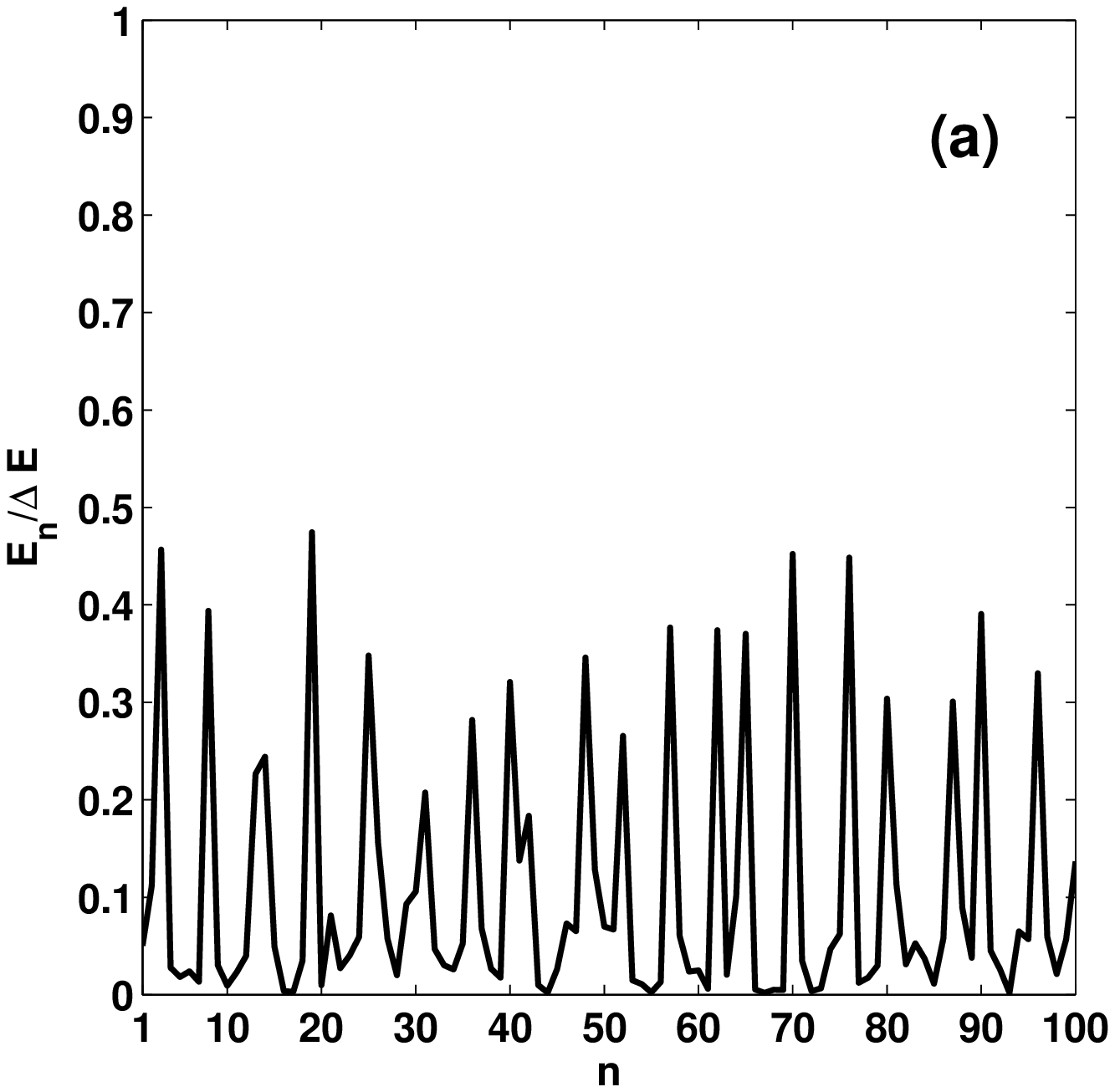}}
 \resizebox{8.cm}{6.cm}{\includegraphics[scale=1.0]{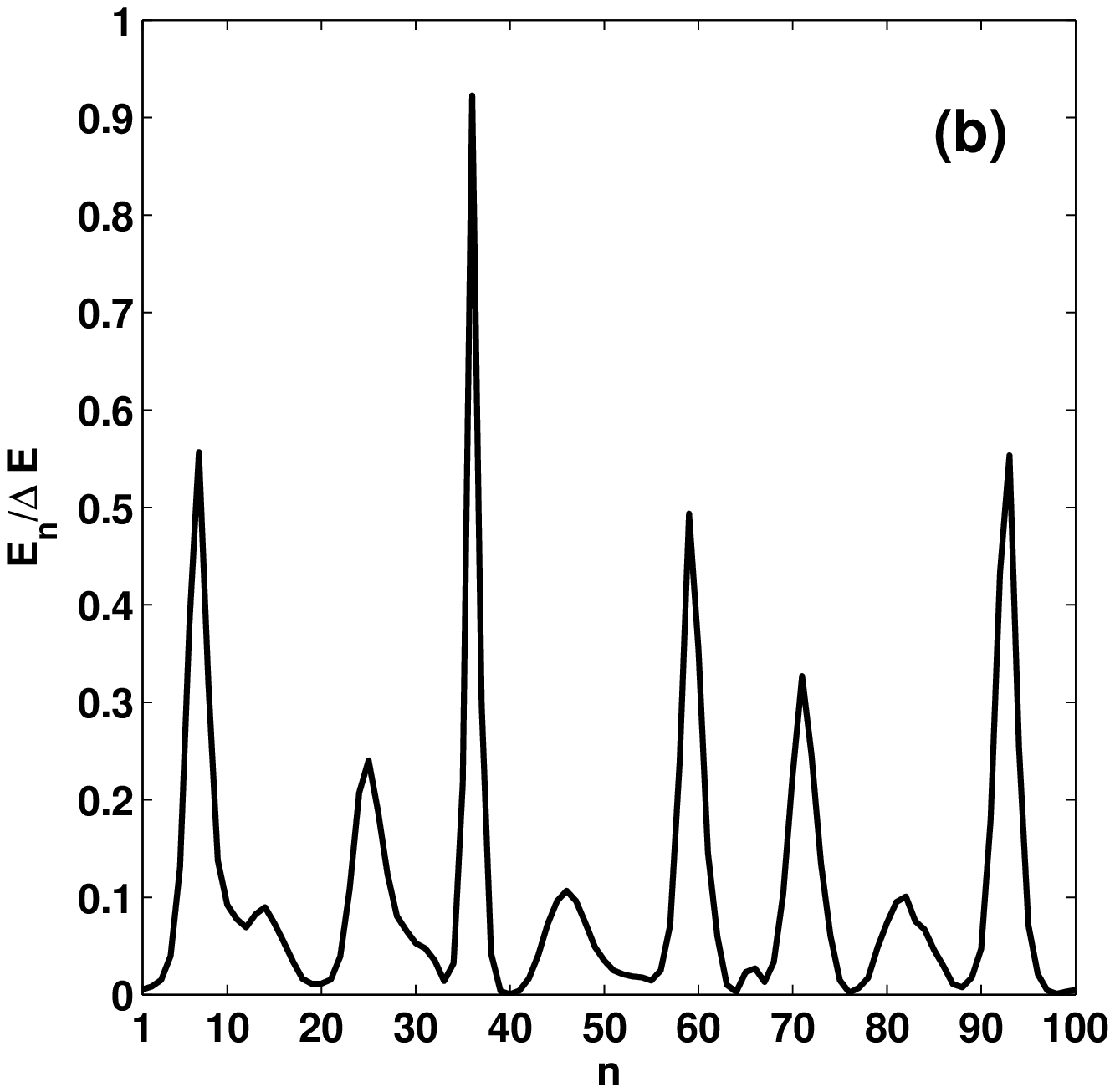}}
\resizebox{8.cm}{6.cm}{\includegraphics[scale=1.0]{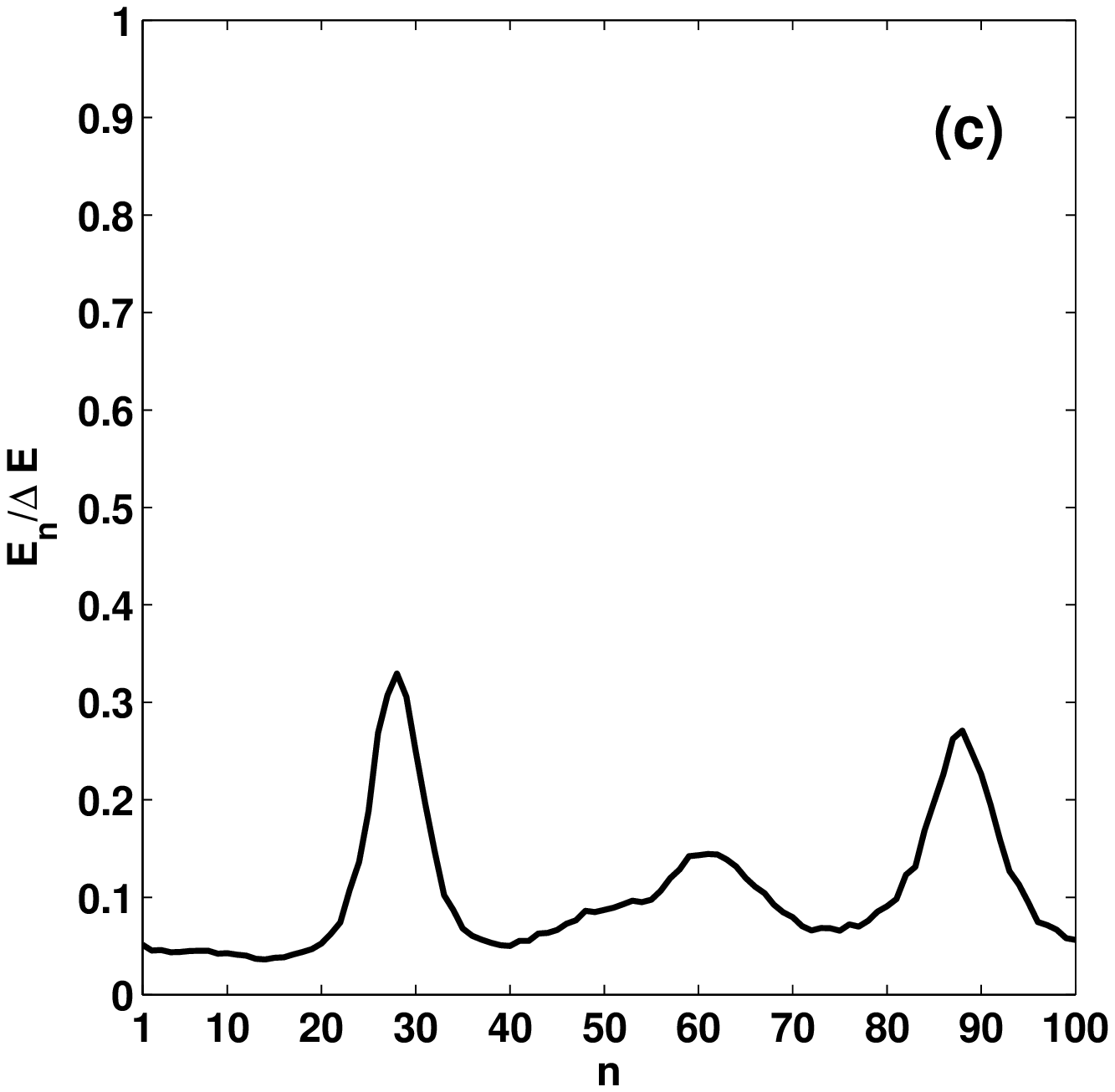}}
    \end{minipage}
    \begin{minipage}{5.cm}
    \end{minipage}
  \end{center}
\caption{\label{fig:pattern}
Spatially localized
structure at $t=500$ for three different coupling strengths. Initial
conditions and parameter values as in Fig.~\ref{fig:density}
except for the coupling strengths
(a) $\kappa=0.05$, (b) $\kappa=0.31$, and (c) $\kappa=1.00$.}
\end{figure}

A physical  explanation for the appearance of a resonance-like
structure can be given in terms of the different degrees of
instability of the underlying motion facilitating the destruction of
the initial flat mode by modulational perturbations. We recollect
that with the variation of the perturbation strength the growth rate
changes (cf. Sec.~\ref{section:modulational}) from a more flat to a
strongly curved single-peaked structure. To illustrate the impact of
the growth rate on the degree of localization of emerging patterns
we present in Fig.~\ref{fig:pattern}
 the energy distribution defined in Eq.~(\ref{eq:endensity})
at an early instant of time---namely, after the formation of the spatially
localized structure due to spontaneous modulational instability has
taken place. For comparison, patterns for three coupling strengths
are shown. In all cases a number of isolated localized humps are
formed. The number of humps, $N_{humps}$, can be attributed to the
wave number at maximal growth rate as follows: $\lambda_{pattern}
N_{humps}=N$ and $\lambda_{pattern}=2\pi/Q_{max}$. Most importantly,
the number of humps (besides their height and width) regulates how
the total energy is shared among them.
Clearly, for $\kappa=0.31$
[Fig.~\ref{fig:pattern}~(b)] the energy is more strongly localized
(fewer humps and of higher height) than in the case of
$\kappa=0.09$ [Fig.~\ref{fig:pattern}~(a)].
In comparison for $\kappa=1$ [Fig.~\ref{fig:pattern}~(c)] the number of humps is further diminished, but they are of lower height
than most of the humps for $\kappa=0.31$.
In fact, for $\kappa=0.31$ the energy contained in the unit at site
$n=35$ is close to the one of the barrier.
Thus, a localized pattern appropriate for escape is provided already
by the mechanism of modulational instability. In particular, no
further (major) energy accumulation, which would delay the escape
process considerably, is hence required.

\begin{figure}
  \begin{center}
    \begin{minipage}{8.cm}
      \resizebox{8.cm}{6.cm}{\includegraphics[scale=1.]{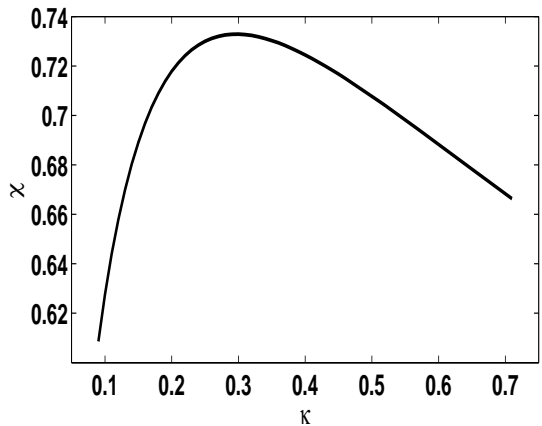}}
    \end{minipage}
    \begin{minipage}{5.cm}
    \end{minipage}
\end{center}
\caption{\label{fig:chi} The ratio $\chi$ defined in
Eq.~(\ref{eq:chiratio}) as a function of the coupling strength
$\kappa$. The parameter values are $\omega_0^2=2$ and  $a=1$. The initial conditions are given by $q_0=0.45$ and $p_0=0$.}
\end{figure}

To gain further insight into the efficiency of energy localization it is illustrative
to suppose that the whole lattice can be divided into a periodic array of
(non-interacting) segments where each of them supports a
single localized hump. The energy of one segment is determined by
\begin{equation}
E_s=\frac{E_{total}}{N_{humps}}=\frac{E_{total}}{N/
\lambda_{pattern}}\,.
\label{eq:Es}
\end{equation}
Defining $e_s=E_s/\Delta E$ as the ratio of the energy per
segment to the net barrier energy we obtain
\begin{equation}
    e_s=\frac{2\pi E_{total}}{Q_{max}N\Delta E}\,.
    \label{eq:epsilon_qmax}
\end{equation}
Appropriate conditions for escape are provided when the energy
contained in each segment, $e_s$, is close to the
activation energy, $e_{act}$, measured in units of the barrier
energy---i.e., $e_{act}=E_{act}/\Delta E$.
The efficiency of energy localization is then determined by the following ratio
\begin{equation}
    \chi=\frac{e_s}{e_{act}}\,.
    \label{eq:chiratio}
\end{equation}
For a given value of the coupling strength $\kappa$ the activation
energy is known; cf. Fig.~\ref{fig:ae}. Fixing the initial energy
and using Eq.~(\ref{eq:growthrate}) we infer the value of $Q_{max}$
and finally using
 Eqs.~(\ref{eq:epsilon_qmax}) and (\ref{eq:chiratio})
we obtain $\chi$. In Fig.~\ref{fig:chi} the ratio $\chi$ is plotted
as a function of the coupling strength $\kappa$. The plot exhibits a
maximum at $\kappa=0.3$, which corroborates the finding of the
resonance found for the escape versus coupling strength as depicted
in Fig.~\ref{fig:resonance}. Moreover, the curvature of the graphs
of Figs.~\ref{fig:resonance} and \ref{fig:chi} are  similar.

\section{Influence of chain length on the escape time}\label{section:length}

We also study the influence of "size"---i.e., the number of
oscillators, $N$, on the escape process. A general constraint on the
escape process arising from varying the chain length is formulated.
We discuss the escape time statistics for chains with constant
energy density and for chains with fixed total energy, but varying
number of oscillators, respectively. Our studies apply to a
homogeneous initial state ($k=0$ mode).

First of all, the initial homogeneous state must become unstable
with respect to a modulational instability. With relation (\ref{eq:constraint_Q})
a constraint is established on the
allowed wave numbers, giving rise to the modulational instability.
They form a discrete set, and we can derive a lower bound for the
number of oscillators, $N_{min}$, needed for the onset of the
modulational instability:
\begin{equation}
    N_{min}>\frac{\pi}{\arcsin\left(\sqrt{\frac{\gamma u_0^2}{2{\kappa}}}\right)}.
    \label{eq:number_Q}
\end{equation}
\noindent
In the case of $u_0^2>2{\kappa}/\gamma$ the initial homogeneous state
is always unstable, independent of the number of oscillators, $N$. However,
for an initial state in the weakly nonlinear regime, which means
$u_0^2< 2{\kappa}/\gamma$, the inequality (\ref{eq:number_Q}) yields
a condition for the minimal number of units on the chain
that are necessary for modulational instability.
On the other hand, once the conditions are provided that the chain be able to adopt the transition state,
the addition of further lattice sites beyond a certain number leaves the activation energy
unaltered. This is due to the fact that the transition state is represented by the c.l.m. which
is strongly localized in space with exponential decaying tails.

\subsection*{Case with constant energy density}
Let us first consider chains with constant energy density
$\rho=E_{total}/N=const$.  One would at first glance expect a faster
escape with increasing number of oscillators and thus with
increasing total energy. This is, however, not necessarily the case.
For an explanation
it is suitable to  consider \textit{the limit of very
long chains}, $N\rightarrow 1000$.

\begin{figure}
  \begin{center}
    \begin{minipage}{8.cm}
      \resizebox{8.cm}{6.cm}{\includegraphics[scale=1.0]{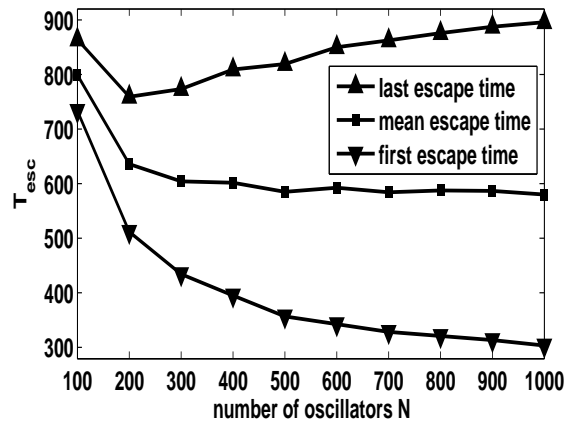}}
    \end{minipage}
    \begin{minipage}{5.cm}
    \end{minipage}
\end{center}
\caption{\label{fig:e_long}First, mean, and last escape time
$T_{esc}$ as a function of the chain length, $N$. The parameter values
are $\omega_0^2=2$, $a=1$, and $\kappa=0.2$. The initial conditions
are given by $q_0=0.50$, $\Delta q=0.01$, and $p_0=\Delta p=0$.}
\end{figure}

In Fig.~\ref{fig:e_long}
the average time for which the {\it first} and {\it last} escape
incidents of a unit take place is depicted versus varying chain
length $N$. The  averages were performed again over $500$
realizations of random initial conditions. In
addition,  we as well depict the mean escape time of the chain. We
set the initial coordinates around a mean value of  $q_0=0.50$ and
spread at  $\Delta q=0.01$,
yielding
$\rho=0.156\Delta E$.
  Apparently the longer the chain is, the more humps (breathers) that are formed due to
   the modulational instability. This offers the possibility that a
   larger number of interacting breathers contribute to an enhanced energy
localization
     in a confined region of the chain which in turn boosts the formation of the critical
      localized mode. Hence, the time it takes for the first unit to escape shrinks
with increasing chain length, while the last escape time increases
due to the enlarged number of escaping units. Also, the mean escape
time
 becomes insensitive to variations of the chain length for sufficiently large
  length $N\geq 500$, and thus it tends to saturate.
\subsection*{Fixed total energy}
\begin{figure}
  \begin{center}
    \begin{minipage}{8.cm}
      \resizebox{8.cm}{6.cm}{\includegraphics[scale=1.]{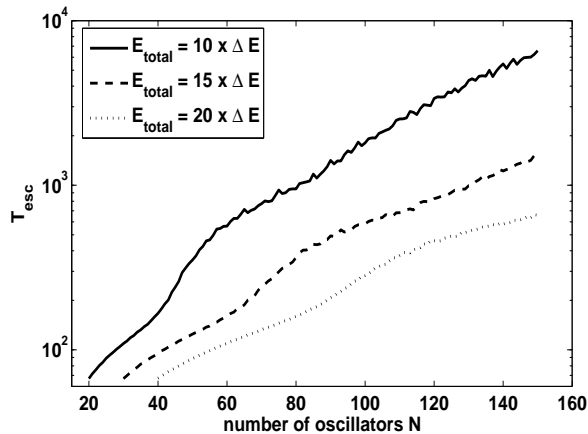}}
    \end{minipage}
    \begin{minipage}{5.cm}
    \end{minipage}
\end{center}
\caption{\label{fig:e_cst}Mean escape time $T_{ecs}$ as a function of the number of oscillators, $N$, for
different values of $E_{total}$. For the choice of $q_0$, see text. Furthermore  $\Delta q=0.01$ and $p_0=\Delta p=0$. The parameter values are $\omega_0^2=2$, $a=1$,
 $\kappa=0.2$, and $E_{total}$ as depicted in the legend.}
\end{figure}

We next consider the situation when a fixed amount of total energy---i.e., $E_{total}=const$---is provided to the system and the number of units on the lattice is varied.
To obtain a certain value of $E_{total}$ upon altering the number
of units, $N$, we adopted $q_0$ appropriately while keeping $\Delta q=0.01$ and $p_0=\Delta p=0$ fixed.
The maximal energy content per unit is restricted to the range
$E_0<0.5\Delta E$.  The results for the mean escape time are
depicted in Fig.~\ref{fig:e_cst}. Generally, we observe an increase
of $T_{esc}$ with growing $N$. Interestingly enough, the slope
passes through intermediate stages of sub-exponential,
hyper-exponential and eventually approaches an exponential behavior.

\section{Summary}

In this paper we have explored the conservative and deterministic
dynamics of a one-dimensional chain consisting of linearly coupled
anharmonic oscillators that are placed into a cubic on-site
potential. Attention has been paid to the collective barrier
crossing of the whole chain. Initially the system is placed into a
metastable state for which all units are trapped near the bottom of
the potential. An overcoming of the barrier of the whole chain is
prevented at short initial times because of a too high net barrier
height. Nevertheless, as we convincingly demonstrated, the
spontaneous formation of localized modes upon evolving time serves
to enrich energetically a segment on the chain to such a degree that
it adopts the transition state energy  in assuming the form of a
hairpin. We have shown that the associated critical localized
lattice state is dynamically unstable and eventually a barrier
crossing proceeds as the propagation of a kink-antikink-like pair
along the chain. 
Strikingly, there exists a resonant-like
coupling strength $\kappa$ for which the escape time (rate) becomes
minimal (maximal), cf. Fig.~\ref{fig:resonance}.



%

In view of potential applications we note that this  deterministic
collective escape process provides  nonlinear systems with the
unique possibility to self-promote their activation dynamics.
Particularly, the ability to operate efficiently---i.e., exhibiting an
enhanced collective coherent escape---although not  optimally
initialized (meaning that one starts out with a far too low energy
density compared to the barrier height) underpins the  beneficial
use of this physical scenario.
Remarkably, while at weak thermal noise the rate of thermal escape
is exponentially suppressed, a deterministic nonlinear breather
dynamics yields a robust critical nucleus configuration, which in
turn  causes an enhancement of the noise-free escape rate. Thus, the
freezing out of noise may prove advantageous for transport in
metastable landscapes, whenever the deterministic escape dynamics
can be launched in a single shot via an initial energy supply.

\vspace{0.5cm}

\centerline{\large{\bf Acknowledgments}}

\noindent This research has been supported by SFB-555 (L.S.-G. and S.F.)
and, as well, by  the joint Volkswagen Foundation Projects No.
I/80424 (P.H.) and No. I/80425 (L.S.-G.).

\end{document}